\documentclass[pdflatex,sn-mathphys-num]{sn-jnl}

\usepackage{graphicx}%
\usepackage{multirow}%
\usepackage{amsmath,amssymb,amsfonts}%
\usepackage{amsthm}%
\usepackage{mathrsfs}%
\usepackage[title]{appendix}%
\usepackage{xcolor}%
\usepackage{textcomp}%
\usepackage{manyfoot}%
\usepackage{booktabs}%
\usepackage{algorithm}%
\usepackage{algorithmicx}%
\usepackage{algpseudocode}%

\theoremstyle{thmstyleone}%
\newtheorem{theorem}{Theorem}%
\newtheorem{proposition}[theorem]{Proposition}%
\newtheorem{lemma}[theorem]{Lemma}%

\theoremstyle{thmstyletwo}%

\theoremstyle{thmstylethree}%
\newtheorem{definition}{Definition}%

\newcommand{\Adv}{\mathrm{Adv}}
\newcommand{\KDF}{\mathrm{KDF}}
\newcommand{\MAC}{\mathrm{MAC}}
\newcommand{\Hash}{\mathrm{H}}
\newcommand{\Kpq}{K_{\mathrm{pq}}}
\newcommand{\Kec}{K_{\mathrm{ec}}}
\newcommand{\Encaps}{\mathsf{Encaps}}
\newcommand{\Decaps}{\mathsf{Decaps}}
\newcommand{\KeyGen}{\mathsf{KeyGen}}
\newcommand{\concat}{\,\|\,}

\begin{document}

\title[Transcript-Bound Combiners for Downgrade-Resilient Hybrid PQ Key Establishment]{Transcript-Bound Combiners for Downgrade-Resilient Hybrid Post-Quantum Key Establishment: Definition, Proof, and Embedded-Device Cost}

\author*[1]{\fnm{Bhanwar} \sur{Gupta}}\email{bgupta55@gmail.com}

\author[1]{\fnm{Sanjeev} \sur{Rana}}\email{sanjeev.rana@mmumullana.org}

\affil*[1]{\orgdiv{Department of Computer and Software Engineering}, \orgname{Maharishi Markandeshwar (Deemed to be University)}, \orgaddress{\city{Mullana, Ambala}, \postcode{133207}, \state{Haryana}, \country{India}}}

\abstract{Hybrid key establishment runs a post-quantum key-encapsulation mechanism (KEM) alongside a classical Diffie--Hellman primitive, so that the session key stays secure while either component resists attack. This design is now standardized in the Transport Layer Security protocol, Secure Shell, and the Internet Key Exchange, with the standardized module-lattice KEM (ML-KEM) as the post-quantum component. A hybrid KEM secures the derived key, but not the integrity of the negotiation that selects which primitives are used. Full protocols authenticate that negotiation through a handshake transcript; a hybrid KEM deployed as a standalone drop-in primitive, or inside a minimal handshake without transcript authentication, inherits no such guarantee, and an active attacker can strip the post-quantum option. We ask what the key schedule alone must contain to make downgrade resilience a local property of the combiner. We give a game-based definition at the combiner layer and prove a two-sided separation: a combiner that ignores the transcript is downgraded with certainty, whereas one that binds the session key and the confirmation tag to a hash of the transcript blocks every such attempt, up to a term negligible for a 256-bit transcript hash. We also give an explicit strongest-link security bound. Using a calibrated cost model composed from published Cortex-M4 measurements, transcript binding adds one hash per party---about 11.8\% of handshake computation but only 1.5\% of radio-inclusive energy---and adds no messages or bytes on the wire. Every reported number is produced by a released harness that passes a 30-check validation gate.}


\maketitle

\section{Introduction}\label{sec:intro}

Shor's algorithm solves the discrete-logarithm and integer-factoring problems efficiently on a large quantum computer~\cite{shor1997}. The public-key primitives underlying today's infrastructure---RSA, elliptic-curve Diffie--Hellman, and DSA---all rest on one of these problems. The timeline for a cryptographically relevant quantum computer is uncertain, but a harvest-now, decrypt-later adversary records encrypted traffic today and decrypts it once such a machine exists. Quantum-resistant key establishment must therefore be deployed before the machine arrives.

NIST's first standardization round concluded in 2024 with ML-KEM (FIPS~203), ML-DSA (FIPS~204), and SLH-DSA (FIPS~205)~\cite{fips203,fips204,fips205,nistir8413}. ML-KEM, descended from CRYSTALS-Kyber~\cite{bos2018kyber}, is the designated key-encapsulation mechanism (KEM). Its security rests on the Module Learning-With-Errors problem, a lattice assumption that is young relative to the decades of cryptanalysis behind elliptic-curve Diffie--Hellman. The classical break of SIKE in 2022~\cite{castryck2023sidh} illustrates how quickly confidence in a new assumption can erode~\cite{bernstein2017postquantum}.

The pragmatic response, already standardized in the Transport Layer Security (TLS) protocol, Secure Shell, and the Internet Key Exchange, is the hybrid KEM. Two peers run a post-quantum KEM and a classical Diffie--Hellman primitive in parallel and derive one session key from both shared secrets. The channel stays secure as long as either primitive survives. Internet-scale experiments with hybrid key exchange~\cite{alkim2016newhope,bos2016frodo} and later TLS performance studies~\cite{paquin2020benchmarking,sikeridis2020pqauth} established that the approach is practical, and its TLS~1.3 design rationale is documented in an IETF draft~\cite{stebila2019hybrid_design}.

\subsection{The negotiation problem}\label{subsec:negproblem}

A hybrid KEM secures the \emph{key} once both parties have agreed on which suite to use. It says nothing about the integrity of the \emph{negotiation} that precedes that agreement. In a configurable deployment, each device advertises the suites it supports and the peers select one. An attacker who controls the network can intercept those advertisements, delete the post-quantum option, and force both peers onto a classical-only handshake that neither would have chosen.

This threat is concrete. The Logjam attack forced TLS peers onto export-grade Diffie--Hellman that the attacker could then factor in real time~\cite{adrian2015logjam}. The analogous move during the quantum transition strips the post-quantum KEM, so two peers that prefer hybrid security complete a purely classical handshake---the exact outcome that motivated deploying ML-KEM.

Standardized full protocols already prevent this. TLS~1.3 folds the negotiated parameters into a hash of the complete handshake transcript and authenticates that hash in a Finished message~\cite{rfc8446}; Bhargavan et al.\ formalized this pattern and proved it yields downgrade resilience at the protocol layer~\cite{bhargavan2016downgrade}. A hybrid KEM embedded in TLS~1.3 inherits that protection. The lightweight authenticated key exchange EDHOC~\cite{rfc9528} likewise secures its cipher-suite negotiation: the initiator's advertised list is processed so that the responder can verify the selected suite is the initiator's most preferred mutually supported one, giving downgrade-protected negotiation even on constrained devices.

\subsection{The gap and this work}\label{subsec:gap}

The gap appears when a hybrid KEM is used \emph{outside} a protocol that binds the negotiation. This is not a corner case. A modern hybrid KEM such as X-Wing~\cite{xwing2024} is deliberately specified as a standalone drop-in primitive with no notion of negotiation, so that it can replace a single KEM anywhere; its own security relies on the surrounding protocol for non-malleability of context. An implementer who pairs two raw KEMs in a bespoke minimal handshake, or who adds a hybrid suite to a custom constrained-device protocol without replicating transcript authentication, obtains a combiner whose key schedule is blind to what was advertised. As we show, a downgrade attack then succeeds with certainty. In such a deployment, downgrade resilience is only as reliable as the correctness of a protocol layer that the combiner cannot see.

We ask a sharp question: what is the minimal change to the hybrid key schedule that makes downgrade resilience a \emph{local} property of the combiner, independent of whether the surrounding protocol authenticates the transcript?

The answer is one hash. Including a hash of the negotiation transcript in the inputs to both the key-derivation function and the key-confirmation code binds the derived session key to the parameters that were actually observed. If an attacker rewrites those parameters, the two parties assemble different transcripts, derive different keys, and the initiator's tag verification fails and aborts. No extra messages, round trips, public-key operations, or wire bytes are introduced.

We are explicit about the size of this claim. The security mechanism is a direct, expected specialization of the transcript-binding principle of Bhargavan et al.~\cite{bhargavan2016downgrade} from the protocol layer to the combiner layer; we do not claim a new cryptographic technique. Our contribution is the combiner-local definition and instantiation, a concrete two-sided bound with an explicit constant for this specific construction, and an engineering-cost account of the mechanism on constrained hardware.

\paragraph{Contributions.}
\begin{enumerate}
  \item A game-based definition of downgrade resilience for hybrid key establishment at the combiner layer, with an active man-in-the-middle adversary, Send oracles, and a matching predicate for the bad event (Section~\ref{sec:model}).
  \item A two-sided separation (Section~\ref{sec:security}): the combiner that ignores the transcript is downgraded with probability~$1$; the transcript-bound combiner has downgrade advantage at most $q_H/2^{n}+\Adv^{\mathrm{euf\text{-}cma}}_{\MAC}$, with $n$ the transcript-hash length. Both sides are confirmed in executed code.
  \item An explicit strongest-link IND-CCA bound for the concrete combiner, with reductions to the component KEMs in the random-oracle model (Section~\ref{sec:security}), corroborated by an exposed-component distinguisher whose success rate tracks the analytic term with $R^2=0.98$.
  \item An engineering account for constrained devices: a constant-time discussion of the added mechanism, and a calibrated cost model composed from published Cortex-M4 per-primitive measurements~\cite{pqm4,lenngren_x25519} with a radio-inclusive energy budget (Sections~\ref{sec:impl} and~\ref{sec:results}).
  \item A worked mapping of the combiner into an EDHOC message flow, and a released harness whose 30-check validation gate must pass before any number is reported (Section~\ref{sec:impl}, Appendices).
\end{enumerate}

\paragraph{Worked motivation.} Consider an initiator $I$ and a responder $R$ that both prefer the hybrid suite \textsf{mlkem768\_x25519} but also advertise classical \textsf{x25519} for interoperability, paired directly through a combiner rather than through TLS or EDHOC. With the plain combiner $K=\KDF(\Kpq\concat\Kec)$, an attacker deletes every post-quantum entry from both advertised lists. $R$ sees only \textsf{x25519}, selects it, and both parties derive a purely classical key, because the key schedule never recorded what was advertised. We reproduce this downgrade at a 100\% rate. Binding the key to a hash of the transcript closes the gap: the transcripts diverge, the keys diverge, and the handshake aborts every time.

\section{Background and Notation}\label{sec:bg}

\subsection{Cryptographic primitives}\label{subsec:prims}

A KEM is a triple $(\KeyGen,\Encaps,\Decaps)$. $\KeyGen$ outputs an encapsulation key $ek$ and a decapsulation key $dk$; $\Encaps(ek)$ outputs a shared secret $K$ and a ciphertext $c$; $\Decaps(dk,c)$ recovers $K$. A KEM is IND-CCA secure if no efficient adversary distinguishes the real shared secret from a random one, even with a decapsulation oracle, except with negligible advantage.

We instantiate the post-quantum component with ML-KEM-768~\cite{fips203}, whose IND-CCA security follows from a Fujisaki--Okamoto-style transform over Module-LWE~\cite{hofheinz2017fo,bos2018kyber}. We treat X25519~\cite{rfc7748,bernstein2006curve25519} as a nominal-group Diffie--Hellman KEM. The key-derivation function $\KDF$ and transcript hash $\Hash$ are SHAKE-256 and SHA3-256~\cite{fips202}; key confirmation uses HMAC-SHA3-256 as the message authentication code $\MAC$~\cite{krawczyk2010hkdf,rfc5869}. Table~\ref{tab:notation} lists all notation. Multi-letter operators such as $\KDF$, $\MAC$, $\Hash$, and $\Adv$ are upright.

\begin{table}[t]
\caption{Notation used throughout the paper.}\label{tab:notation}
\begin{tabular}{@{}ll@{}}
\toprule
Symbol & Meaning \\
\midrule
$I,\,R$ & Initiator and responder \\
$\mathcal{A}$ & Active man-in-the-middle adversary \\
$\Kpq,\,\Kec$ & ML-KEM-768 and X25519 shared secrets \\
$ek,\,dk,\,c$ & Encapsulation key, decapsulation key, ciphertext \\
$A_I,\,A_R$ & Advertised cipher-suite lists \\
$\tau$ & Handshake transcript as observed by one party \\
$\ell$ & Domain-separation label \\
$K$ & Derived session key \\
$\Hash,\,\KDF,\,\MAC$ & Hash, key-derivation function, message authentication code \\
$q_H$ & Adversary's random-oracle query budget \\
$n$ & Transcript-hash output length in bits \\
$\gamma$ & $\min(|\Kpq|,|\Kec|)$ in bits \\
$\Adv^{X}(\mathcal{A})$ & Advantage of $\mathcal{A}$ in game $X$ \\
\botrule
\end{tabular}
\end{table}

\subsection{Hybrid combiners}\label{subsec:combiners}

A combiner turns two component shared secrets into one session key. The two combiners we compare share an identical protocol skeleton and differ only in the key schedule:
\begin{align}
\text{Plain:}\quad & K = \KDF(\Kpq \concat \Kec), \label{eq:naive}\\
\text{Bound:}\quad & K = \KDF(\ell \concat \Kpq \concat \Kec \concat \Hash(\tau)), \label{eq:bound}
\end{align}
where $\tau$ is the ordered concatenation of every handshake message as the local party observed it, including both advertised suite lists and the selected suite. The confirmation tag from the responder is $\MAC(K,\texttt{"finished"})$ for the plain variant and $\MAC(K,\tau)$ for the bound variant. Equation~\eqref{eq:bound} is the only change we advocate.

The intuition is direct. In the plain combiner, $K$ is a function of only the two shared secrets, so rewriting the advertised lists leaves no trace in $K$: the two parties derive the same key even though they negotiated on different lists. In the bound combiner, $K$ also depends on $\Hash(\tau)$. If an attacker alters the lists, the parties assemble different transcripts $\tau_I\neq\tau_R$, derive different keys $K_I\neq K_R$ except with negligible probability, and the initiator's verification of the responder's tag fails.

\section{Related Work}\label{sec:related}

\paragraph{KEM combiners and hybrid KEMs.} Giacon, Heuer, and Poettering formalized KEM combiners and proved that a split-key pseudorandom-function core achieves IND-CCA security whenever one component KEM is secure~\cite{giacon2018kem}. Bindel et al.\ extended hybrid guarantees to authenticated key exchange and to quantum adversaries in the quantum random-oracle model (QROM)~\cite{bindel2019hybrid}; Huguenin-Dumittan and Vaudenay analysed Fujisaki--Okamoto-style combiners in the same setting~\cite{huguenin2021folike}. The closest concrete design is X-Wing~\cite{xwing2024}, which fixes ML-KEM-768 and X25519 and proves IND-CCA security using SHA3-256. X-Wing is a standalone KEM: it omits negotiation by design so it can serve as a drop-in primitive, and its specification notes that non-malleability of surrounding context is provided by the embedding protocol. These works establish the strongest-link guarantee our combiner inherits; none models suite negotiation or defines downgrade resilience at the combiner layer.

\paragraph{Hybrid authenticated key exchange.} Dowling, Hansen, and Paterson give a framework for provably quantum-secure hybrid authenticated key exchange with whole-protocol security against component compromise~\cite{dowling2020muckle}. Our scope is narrower by design: a combiner-level property, its proof, and its cost, rather than a full authenticated key exchange model.

\paragraph{Downgrade resilience.} Bhargavan et al.\ introduced downgrade resilience as a formal notion for key-exchange protocols and identified the transcript-binding patterns that guarantee it~\cite{bhargavan2016downgrade}; TLS~1.3~\cite{rfc8446} and EDHOC~\cite{rfc9528} adopt those patterns. Logjam is the canonical real-world downgrade attack~\cite{adrian2015logjam}. We apply the same principle inside the combiner's key schedule, so that the guarantee holds even when the embedding protocol does not provide it, and we give a combiner-specific quantitative bound.

\paragraph{Post-quantum and constrained key exchange.} KEMTLS replaces handshake signatures with KEMs~\cite{schwabe2020kemtls}. Hybrid and post-quantum TLS and SSH have been prototyped and benchmarked~\cite{crockett2019prototyping,paquin2020benchmarking,sikeridis2020pqauth}. For constrained nodes~\cite{rfc7228}, EDHOC provides lightweight authenticated Diffie--Hellman over CoAP and OSCORE~\cite{rfc9528,rfc7252,rfc8613}.

\paragraph{Embedded and physical-security engineering of ML-KEM.} Because our cost claims target constrained devices, we position the mechanism against the embedded ML-KEM literature. The pqm4 framework is the standard reference for Cortex-M4 cycle and memory measurements of ML-KEM~\cite{pqm4}, and Lenngren's assembly implementation is the standard optimized X25519 for the same core~\cite{lenngren_x25519}. Side-channel resistance for lattice KEMs is an active area: masking Kyber at first and higher orders~\cite{bos2021maskingkyber}, first-order masked Kyber on Cortex-M4~\cite{heinz2022maskedkyber}, efficient masking-conversion techniques~\cite{bronchain2022bitslice}, and demonstrated side-channel attacks on masked lattice KEMs~\cite{ngo2021sidechannel}. This body of work matters here for two reasons: it fixes the per-primitive costs our model composes, and it sets the scale against which the binding overhead should be read---a masked ML-KEM-768 decapsulation costs roughly an order of magnitude more than one unmasked decapsulation, so a single transcript hash is negligible in any side-channel-hardened deployment (Section~\ref{sec:impl}).

Table~\ref{tab:related} positions this work against the closest prior contributions. A check mark in \emph{SL} means the work proves a strongest-link IND-CCA guarantee; \emph{DR} means it establishes downgrade resilience; \emph{Neg} means it explicitly models suite negotiation; \emph{M4} means it reports Cortex-M4 cost data for the construction. Our contribution is the row that combines all four.

\begin{table}[t]
\caption{Comparison with the closest prior work. SL: strongest-link IND-CCA; DR: downgrade resilience; Neg: suite negotiation modelled; M4: Cortex-M4 cost data.}\label{tab:related}
\begin{tabular}{@{}lcccc@{}}
\toprule
Work & SL & DR & Neg & M4 \\
\midrule
Giacon et al.~\cite{giacon2018kem}       & \checkmark & & & \\
Bindel et al.~\cite{bindel2019hybrid}    & \checkmark & & & \\
X-Wing~\cite{xwing2024}                  & \checkmark & & & \\
Dowling et al.~\cite{dowling2020muckle}  & \checkmark & & & \\
Bhargavan et al.~\cite{bhargavan2016downgrade} & & \checkmark & \checkmark & \\
This work                                & \checkmark & \checkmark & \checkmark & \checkmark \\
\botrule
\end{tabular}
\end{table}

\section{Problem and Threat Model}\label{sec:model}

\subsection{Setting}\label{subsec:setting}

Two honest parties, initiator $I$ and responder $R$, each hold a local configuration listing the cipher suites they support, drawn from $\{\textsf{mlkem768\_x25519},\,\textsf{mlkem768},\,\textsf{x25519}\}$ and ranked by a public strength order with the hybrid suite strongest. They negotiate the strongest mutually supported suite. The message flow mirrors hybrid key-share negotiation in TLS~1.3, IKEv2, and lightweight constrained key exchange: $I$ sends its advertised list and key shares; $R$ selects the strongest mutual suite, performs the KEM or Diffie--Hellman operation, and returns its list, selection, shares, and a confirmation tag; $I$ completes and verifies the tag. Entity authentication is assumed to be provided by the embedding protocol or by pre-shared credentials; our concern is the integrity of the negotiated parameters.

\subsection{Adversary}\label{subsec:adv}

The adversary $\mathcal{A}$ is an active man-in-the-middle with full network control. It may read, drop, reorder, and rewrite any handshake message, including the advertised lists and the selected suite. It does not break the underlying primitives, and it makes at most $q_H$ random-oracle queries. This is the standard downgrade setting of Bhargavan et al.~\cite{bhargavan2016downgrade}.

\begin{definition}[Downgrade resilience]\label{def:dr}
Consider the game $\mathbf{G}^{\mathrm{dr}}_{\Pi}$ in which a challenger initializes honest parties $I$ and $R$ whose configurations share a strongest mutual suite $s^\star$, and runs one handshake with all messages routed through $\mathcal{A}$ via $\mathsf{Send}$ oracles that deliver $\mathcal{A}$'s chosen bitstrings. $\mathcal{A}$ wins if some honest session completes on a suite weaker than $s^\star$. The advantage is $\Adv^{\mathrm{dr}}_{\Pi}(\mathcal{A})=\Pr[\mathcal{A}\text{ wins }\mathbf{G}^{\mathrm{dr}}_{\Pi}]$. $\Pi$ is downgrade-resilient if this advantage is negligible for every efficient $\mathcal{A}$.
\end{definition}

The bad event is an honest session accepting a key derived from fewer or weaker primitives than both parties were willing to support. This captures an attacker stripping the post-quantum option and forcing a classical-only handshake between two peers that both preferred hybrid security.

\section{The Transcript-Bound Combiner}\label{sec:method}

The construction is Equation~\eqref{eq:bound}, specified in Algorithm~\ref{alg:proto}. The key design decision is that each party assembles $\tau$ from the messages exactly as it received them: $I$ includes its own advertised list and what it received from $R$; $R$ includes what it received from $I$ together with its own list and its selection. Both the session key and the confirmation tag are bound to $\Hash(\tau)$.

If $\mathcal{A}$ rewrites the negotiation---for instance by deleting post-quantum entries---the parties assemble different transcripts $\tau_I\neq\tau_R$. Because $\Hash(\tau)$ enters the $\KDF$ input, the two keys $K_I$ and $K_R$ differ except with negligible probability, so the initiator's check $\MAC(K_I,\tau_I)=t_R$ fails and the handshake aborts. An attacker who leaves the transcript intact cannot downgrade: the parties agree on $s^\star$ and derive $K$ from the hybrid primitive.

The label $\ell$ provides domain separation between combiners built on the same primitives, and the length-prefixed encoding makes the $\KDF$ input injective over distinct component secrets. The added cost over the plain combiner is one evaluation of $\Hash$ over $\tau$ per party; no round trips, public-key operations, messages, or wire bytes are added (Proposition~\ref{prop:cost}).

\begin{algorithm}[t]
\caption{Transcript-bound hybrid key establishment. The plain combiner replaces lines 7 and 12 with $K=\KDF(\Kpq\concat\Kec)$ and uses a constant string as the confirmation input.}\label{alg:proto}
\begin{algorithmic}[1]
\State \textbf{Initiator} $I$: $(ek,dk)\gets\KeyGen_{\mathrm{pq}}$;\; $(sk_I,pk_I)\gets\KeyGen_{\mathrm{ec}}$
\State $I\to R$: $A_I,\; ek,\; pk_I$
\State \textbf{Responder} $R$: $s\gets$ strongest mutual suite from $(A_I,A_R)$
\State \quad if $s$ uses pq: $(\Kpq,c)\gets\Encaps(ek)$
\State \quad if $s$ uses ec: $(sk_R,pk_R)\gets\KeyGen_{\mathrm{ec}}$;\; $\Kec\gets sk_R\cdot pk_I$
\State \quad $\tau_R\gets(A_I,A_R,s,ek,c,pk_I,pk_R)$
\State \quad $K_R\gets\KDF(\ell\concat\Kpq\concat\Kec\concat\Hash(\tau_R))$;\; $t_R\gets\MAC(K_R,\tau_R)$
\State $R\to I$: $A_R,\; s,\; c,\; pk_R,\; t_R$
\State \textbf{Initiator} $I$: if $s$ uses pq: $\Kpq\gets\Decaps(dk,c)$
\State \quad if $s$ uses ec: $\Kec\gets sk_I\cdot pk_R$
\State \quad $\tau_I\gets(A_I,A_R,s,ek,c,pk_I,pk_R)$
\State \quad $K_I\gets\KDF(\ell\concat\Kpq\concat\Kec\concat\Hash(\tau_I))$
\State \quad \textbf{if} $\MAC(K_I,\tau_I)\neq t_R$ \textbf{then abort else} accept $K_I$
\end{algorithmic}
\end{algorithm}

\section{Security Analysis}\label{sec:security}

We work in the random-oracle model with $\KDF$ and $\Hash$ modeled as random oracles; component KEM advantages are in the standard IND-CCA sense. Full proofs are in Appendix~\ref{secA1}; a quantum-random-oracle sketch is in Appendix~\ref{secA2}.

\begin{lemma}[Correctness]\label{lem:correct}
In an honest, unmodified handshake on the hybrid suite, $I$ and $R$ derive the same session key and $I$ accepts, except with probability at most $\delta_{\mathrm{MLKEM}}\approx 2^{-164}$, the ML-KEM-768 decapsulation-failure probability~\cite{fips203}.
\end{lemma}

When no message is rewritten, $\tau_I=\tau_R$, so both inputs to $\KDF$ coincide and the tag check passes; the only failure mode is ML-KEM decapsulation failure.

\begin{theorem}[Strongest-link IND-CCA]\label{thm:strong}
Let the hybrid KEM use the bound combiner of Equation~\eqref{eq:bound} with $\KDF$ a random oracle. For any efficient adversary $\mathcal{A}$ making at most $q_H$ random-oracle queries,
\[
  \Adv^{\mathrm{ind\text{-}cca}}_{\mathrm{hyb}}(\mathcal{A})
  \le
  \min\!\bigl(\Adv^{\mathrm{ind\text{-}cca}}_{\mathrm{pq}}(\mathcal{B}_{\mathrm{pq}}),\,
             \Adv^{\mathrm{ind\text{-}cca}}_{\mathrm{ec}}(\mathcal{B}_{\mathrm{ec}})\bigr)
  + \frac{q_H}{2^{\gamma}},
\]
where $\gamma=\min(|\Kpq|,|\Kec|)$ in bits, and $\mathcal{B}_{\mathrm{pq}}$, $\mathcal{B}_{\mathrm{ec}}$ are explicit reductions running in essentially the same time as $\mathcal{A}$.
\end{theorem}

The $\min$ form is deliberately conservative: breaking the hybrid requires recovering at least one component secret, so the advantage is bounded by the advantage against the harder component. Independence would give the tighter product $\Adv_{\mathrm{pq}}\cdot\Adv_{\mathrm{ec}}$, which we do not assume. This is the split-key argument of Giacon et al.~\cite{giacon2018kem}; our contribution is the explicit $q_H/2^{\gamma}$ constant for this concrete combiner, together with its empirical corroboration in Section~\ref{sec:results}. Bindel et al.~\cite{bindel2019hybrid} prove that hash-based combiners of this form retain the strongest-link guarantee in the QROM, with bounds degraded by the standard square-root factor from Grover search; the construction inherits that qualitative QROM security, and Appendix~\ref{secA2} sketches the lifting argument. A tight QROM constant for the explicit bound is left to future work.

\begin{theorem}[Two-sided downgrade resilience]\label{thm:dr}
Against the adversary of Definition~\ref{def:dr}:
\begin{enumerate}
  \item[(i)] for the plain combiner, there exists an adversary achieving $\Adv^{\mathrm{dr}}=1$;
  \item[(ii)] for the bound combiner, $\Adv^{\mathrm{dr}}\le q_H/2^{n}+\Adv^{\mathrm{euf\text{-}cma}}_{\MAC}$, where $n$ is the transcript-hash output length.
\end{enumerate}
\end{theorem}

With $n=256$ (SHA3-256), the dominant term is $q_H/2^{256}<2^{-190}$ for any realistic query budget. The result is information-theoretic up to the collision and forgery terms: it holds against computationally unbounded adversaries, and for quantum adversaries $2^n$ should be read as $2^{n/2}$ in the worst case.

\begin{proposition}[Binding cost]\label{prop:cost}
Compared to the plain combiner, the bound combiner adds exactly one evaluation of $\Hash$ over $\tau$ per party, and no additional public-key operations, messages, round trips, or transmitted bytes.
\end{proposition}

\section{Implementation and Engineering Considerations}\label{sec:impl}

\subsection{Reference implementation and toolchain}\label{subsec:impl}

We provide a reference implementation of both combiners that executes the full protocol of Algorithm~\ref{alg:proto}. The post-quantum component is ML-KEM-768 through a reference implementation of FIPS~203~\cite{kyberpy}; X25519 is a vetted constant-time library; and the symmetric primitives are the SHA3-family functions of Section~\ref{subsec:prims}. The two combiners share one code path and differ only in the key-schedule and confirmation-tag inputs, so that any measured difference is attributable to transcript binding alone. All functional experiments---correctness, downgrade, and the exposed-component distinguisher---run against this implementation with fixed, logged seeds.

\subsection{Constant-time and side-channel considerations}\label{subsec:sca}

Because the target is constrained devices, we state the side-channel surface of the added mechanism explicitly. The transcript $\tau$ is public: it consists of advertised suite lists, the selected suite, and the public keys and ciphertext already sent on the wire. Computing $\Hash(\tau)$ therefore processes only attacker-observable data and introduces no secret-dependent branch or memory access, so it adds no confidentiality-relevant leakage surface over the plain combiner. The key-confirmation check compares MAC tags; our implementation uses a constant-time comparison, so tag verification does not leak the number of matching bytes. The session-key derivation mixes the transcript hash with the two secret shared values inside a single $\KDF$ call whose leakage profile is that of the underlying SHA3 permutation, unchanged by the added public input.

The confidentiality-critical leakage in this construction is entirely inside the component primitives, not in the combiner. Any deployment that requires side-channel resistance must use a hardened ML-KEM and a constant-time X25519; the relevant literature includes masked Kyber at first and higher orders~\cite{bos2021maskingkyber,heinz2022maskedkyber}, masking-conversion techniques~\cite{bronchain2022bitslice}, and attacks that motivate them~\cite{ngo2021sidechannel}. This sets the scale for the binding overhead: a first-order masked ML-KEM-768 decapsulation on Cortex-M4 costs roughly $3.0$~million cycles~\cite{heinz2022maskedkyber}, about an order of magnitude more than one transcript hash over the full negotiation. Transcript binding is thus negligible precisely in the hardened deployments where side-channel cost dominates.

\subsection{Cost-model methodology}\label{subsec:method}

We are explicit that the device-cost figures in Section~\ref{sec:results} are a calibrated analytic model, not measurements taken on a single instrumented board running the whole protocol. The model composes published per-primitive Cortex-M4F measurements: ML-KEM-768 cycle counts and peak stack from the pqm4 \texttt{m4fspeed} benchmarks on an STM32F4-class core~\cite{pqm4}, and X25519 scalar-multiplication cost from Lenngren's optimized Cortex-M4 implementation~\cite{lenngren_x25519}. The transcript-hash cost is derived from the optimized Keccak-$f$ permutation count (12{,}969 cycles) used by ML-KEM's own hashing on the same core, applied to the number of SHA3-256 blocks spanned by $\tau$. Compute energy uses an STM32L4 operating point (80~MHz, 33~mW); radio energy uses an IEEE~802.15.4-class model (4.8~$\mu$J per byte); the CR2032 budget uses a nominal 225~mAh at 3~V with 80\% usable capacity. Because the two cycle inputs come from independent measurement campaigns on comparable Cortex-M4F cores, absolute figures should be read as a calibrated model; the relative comparisons that carry the argument---binding versus baseline, compute versus radio---are platform-independent. Reproducing these figures on a single instrumented board is the natural next step and is the main threat to external validity (Section~\ref{sec:disc}).

\subsection{Mapping onto a lightweight protocol}\label{subsec:edhoc}

To make the constrained-device framing concrete, we sketch how the transcript-bound combiner composes with EDHOC~\cite{rfc9528} without altering EDHOC's own guarantees. EDHOC already carries the initiator's ordered suite list \textsf{SUITES\_I} and maintains running transcript hashes across messages. A hybrid EDHOC cipher suite would name (ML-KEM-768, X25519) as its key-exchange components; the combiner of Equation~\eqref{eq:bound} is invoked when both components' shared secrets are available, with $\tau$ instantiated from the EDHOC transcript hash that already covers \textsf{SUITES\_I} and the selected suite. In this composition the combiner's binding is defence-in-depth: EDHOC's transcript already secures the negotiation, so the two mechanisms agree. The value of combiner-layer binding is realized in the complementary case---a bespoke minimal handshake or a standalone drop-in KEM pairing with no such transcript---where the combiner is the only layer that authenticates the negotiation. The property is thus local to the combiner and does not depend on the embedding protocol being present or correctly configured.

\subsection{Claim taxonomy and validation gate}\label{subsec:gate}

Table~\ref{tab:claims} classifies every quantitative claim as Proven (P), Measured/derived from published inputs (M), or Future (F). Before any number is reported, a 30-check validation gate must pass; its categories are itemized in Appendix~\ref{secA3} (Table~\ref{tab:gate}). In the released run the gate passes all 30 checks. All experiments use fixed seeds, logged in the artifact.

\begin{table}[t]
\caption{Claim taxonomy. P: proven; M: measured or derived from published inputs; F: future work.}\label{tab:claims}
\begin{tabular}{@{}llc@{}}
\toprule
Claim & Basis & Type \\
\midrule
Plain combiner downgraded w.p.\ $1$            & Thm.~\ref{thm:dr}(i), E2  & P/M \\
Bound combiner aborts every downgrade          & Thm.~\ref{thm:dr}(ii), E2 & P/M \\
Strongest-link bound with $q_H/2^{\gamma}$ term & Thm.~\ref{thm:strong}, E3 & P/M \\
Binding $=11.8\%$ compute, $1.5\%$ energy      & cost model, E5, E6         & M   \\
Peak stack $\approx 6.5$~KB                    & pqm4, E5                   & M   \\
$\approx 80$ bound-hybrid handshakes/day       & E6                         & M   \\
Two latency regimes                            & E7                         & M   \\
Tight QROM constant                            & ---                        & F   \\
\botrule
\end{tabular}
\end{table}

\section{Results}\label{sec:results}

\begin{figure}[t]
\centering
\includegraphics[width=\textwidth]{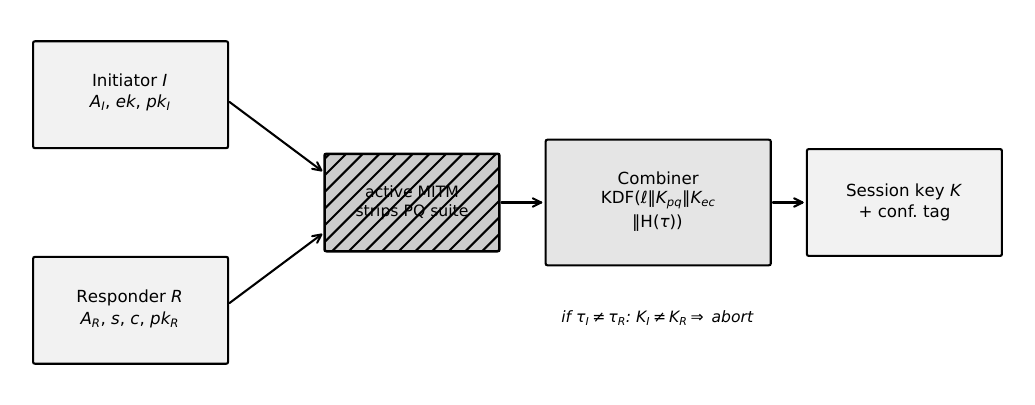}
\caption{Transcript-bound hybrid key establishment. The combiner mixes both component secrets and the transcript hash; an active man-in-the-middle who strips the post-quantum suite changes $\tau$, causing $K_I\neq K_R$ and forcing an abort}\label{fig:arch}
\end{figure}

Figure~\ref{fig:arch} summarizes the construction and the attack surface.

\subsection{Correctness and downgrade resilience (E1, E2)}\label{subsec:e12}

\begin{figure}[t]
\centering
\includegraphics[width=\columnwidth]{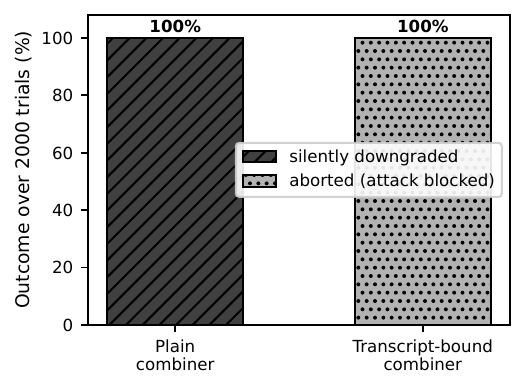}
\caption{Downgrade outcome across 2{,}000 trials. The plain combiner is silently degraded to the classical suite in every trial; the transcript-bound combiner aborts every attempt}\label{fig:downgrade}
\end{figure}

Over 2{,}000 honest hybrid handshakes, $I$ and $R$ agreed on the session key with zero mismatches, consistent with Lemma~\ref{lem:correct}. Under the active adversary (Figure~\ref{fig:downgrade}), the plain combiner was silently downgraded to classical \textsf{x25519} in every trial, while the bound combiner aborted in every trial. Both outcomes are deterministic under the stated adversary, as Theorem~\ref{thm:dr} predicts; the repetition over independently sampled KEM and Diffie--Hellman keys serves to rule out implementation non-determinism rather than to estimate a probability, and no variance is expected or observed.

\subsection{Strongest-link distinguisher (E3)}\label{subsec:e3}

\begin{figure}[t]
\centering
\includegraphics[width=\columnwidth]{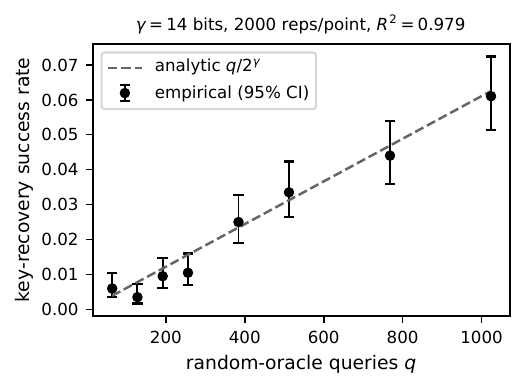}
\caption{Exposed-component distinguisher with 95\% confidence intervals. With one component fully broken, key-recovery success grows as $q/2^{\gamma}$; the fitted slope tracks the analytic line ($R^2=0.98$) and extrapolates below $2^{-190}$ at full parameters}\label{fig:dist}
\end{figure}

To verify the $q_H/2^{\gamma}$ term in Theorem~\ref{thm:strong} empirically, we ran an exposed-component distinguisher: one component secret is revealed in full, and the adversary makes $q$ random guesses of the other over a reduced $\gamma=14$-bit space, winning if any guess reproduces the session key. Figure~\ref{fig:dist} shows the empirical success rate against the analytic $q/2^{\gamma}$ line, with 95\% Wilson intervals over 2{,}000 independent seeds per point. A least-squares fit through the origin has slope $5.91\times10^{-5}$ against the analytic $6.10\times10^{-5}$ (ratio $0.97$) with $R^2=0.98$; every analytic value lies within the empirical confidence interval. Extrapolated to the full $\gamma=256$ bits with $q_H=2^{64}$, the success probability is below $2^{-190}$. As construction sanity checks, $\KDF$ outputs were statistically uniform (monobit fraction $0.499$) when one secret was known, and the SHA3-256 known-answer test passed.

\subsection{Computation and communication (E5)}\label{subsec:e5}

\begin{figure}[t]
\centering
\includegraphics[width=\columnwidth]{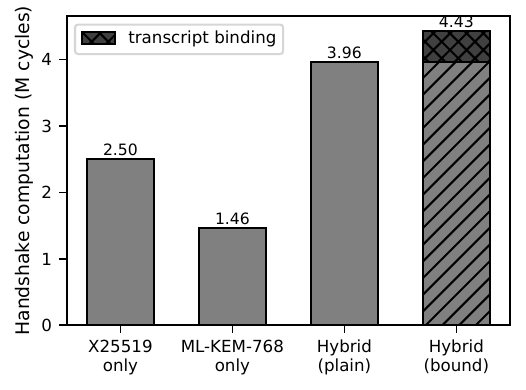}
\caption{Handshake computation from composed Cortex-M4 per-primitive measurements (pqm4 ML-KEM-768; Lenngren X25519). Transcript binding adds one hash evaluation per party}\label{fig:cycles}
\end{figure}

\begin{figure}[t]
\centering
\includegraphics[width=\columnwidth]{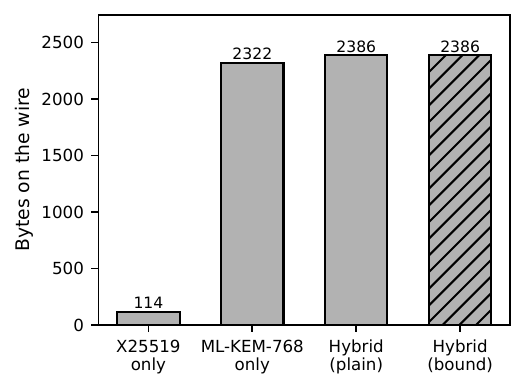}
\caption{Bytes on the wire. The transcript-bound and plain hybrid handshakes are identical in communication cost; binding adds no transmitted bytes}\label{fig:bytes}
\end{figure}

Figure~\ref{fig:cycles} reports handshake computation from the cost model of Section~\ref{subsec:method}. The bound hybrid totals 4.43~M cycles across both parties, against 3.96~M for the plain hybrid; transcript binding adds 0.47~M cycles, or 11.8\%. The hybrid handshake places 2{,}386 bytes on the wire (Figure~\ref{fig:bytes}); binding adds none. Peak stack is 6{,}468 bytes, dominated by ML-KEM-768 and within the Class-2 constrained-node memory budget (Table~\ref{tab:cost}). The model is internally consistent: the plain-hybrid cycle count is exactly the sum of the ML-KEM-only and X25519-only handshakes, since the hybrid performs the union of their operations.

\begin{table}[t]
\caption{Per-handshake cost (both parties combined). Cycle counts and peak stack composed from published Cortex-M4 inputs; energy is split into compute and 802.15.4-class radio}\label{tab:cost}
\begin{tabular}{@{}lrrrr@{}}
\toprule
Scheme & Bytes & Mcyc & Stack (B) & Energy (mJ) \\
\midrule
X25519 only        & 114  & 2.50 & 1000 & 1.58 \\
ML-KEM-768 only    & 2322 & 1.46 & 6468 & 11.75 \\
Hybrid (plain)     & 2386 & 3.96 & 6468 & 13.09 \\
Hybrid (bound)     & 2386 & 4.43 & 6468 & 13.28 \\
\botrule
\end{tabular}
\end{table}

\subsection{Energy (E6)}\label{subsec:e6}

\begin{figure}[t]
\centering
\includegraphics[width=\columnwidth]{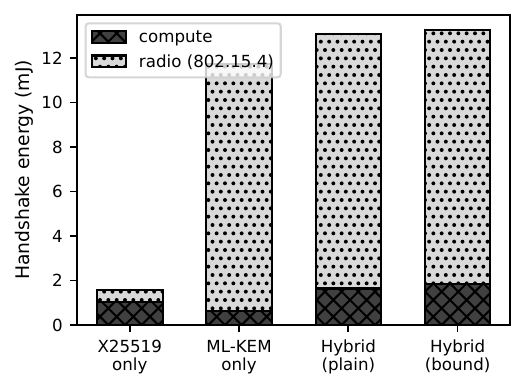}
\caption{Total handshake energy split by compute and radio. Radio transmission dominates; the transcript-binding overhead is a small fraction of total energy}\label{fig:energy}
\end{figure}

\begin{figure}[t]
\centering
\includegraphics[width=\columnwidth]{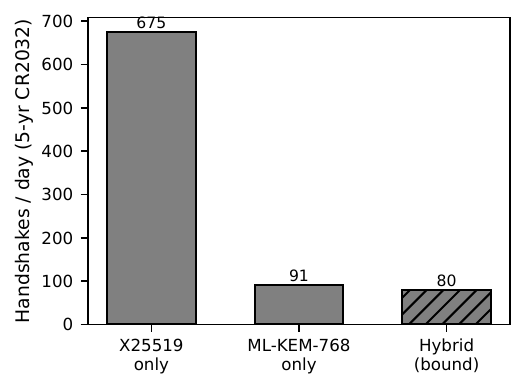}
\caption{Handshakes per day on a five-year CR2032 budget, radio-inclusive. The hybrid budget is dominated by the ML-KEM-768 payload, not by transcript binding}\label{fig:scale}
\end{figure}

Figure~\ref{fig:energy} separates compute from radio energy. For the bound hybrid, compute accounts for 1.83~mJ and radio for 11.45~mJ, making radio 86\% of the 13.28~mJ total. Transcript binding is 11.8\% of compute energy but only 1.5\% of the total handshake budget. The dominant post-quantum cost on a constrained radio link is transmitting the larger ML-KEM-768 key material and ciphertext, not the binding hash. On a five-year CR2032 budget, the radio-inclusive energy supports about 80 bound-hybrid handshakes per day, against 91 for ML-KEM-768 alone and 675 for X25519 alone (Figure~\ref{fig:scale}); the hybrid and ML-KEM-only figures are close because both are dominated by the ML-KEM-768 payload.

\subsection{Latency and security level (E7)}\label{subsec:e7}

\begin{figure}[t]
\centering
\includegraphics[width=\columnwidth]{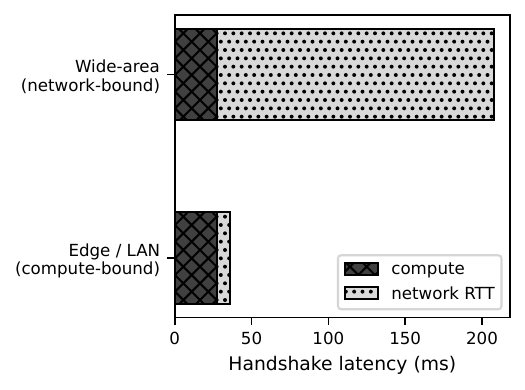}
\caption{Latency projection over published round-trip datasets. Wide-area handshakes are network-bound; edge handshakes are compute-bound}\label{fig:rtt}
\end{figure}

\begin{figure}[t]
\centering
\includegraphics[width=\columnwidth]{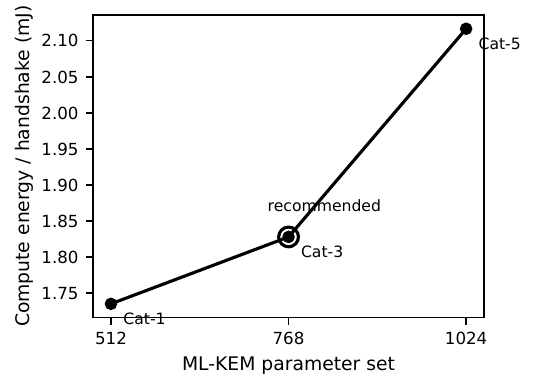}
\caption{Security level versus per-handshake compute energy across ML-KEM parameter sets. ML-KEM-768 is the recommended operating point}\label{fig:knee}
\end{figure}

A latency projection over published round-trip datasets~\cite{netlatency_data} shows two regimes (Figure~\ref{fig:rtt}): wide-area handshakes are network-bound (about 208~ms total), while edge handshakes are compute-bound and cluster near 36~ms. These are projections over real latency data, not measurements of a deployed system. Across ML-KEM security levels (Figure~\ref{fig:knee}), level 768 sits at the efficiency knee: it provides NIST Category-3 security at modest incremental compute energy over level 512, while level 1024 offers a margin most constrained deployments do not require.

\section{Discussion}\label{sec:disc}

\paragraph{Why the defence works.} The construction reduces to one invariant: the session key is a function of the negotiation transcript, so any party that accepts has committed to exactly what it observed on the wire. Theorem~\ref{thm:dr}(ii) makes this precise, and the empirical result follows from the theorem rather than from an implementation detail. The plain combiner fails for the complementary reason: its key schedule is blind to the advertised lists, so an attacker who rewrites them leaves no trace in the key. This is the transcript-authentication principle of TLS~1.3 and EDHOC~\cite{rfc8446,rfc9528,bhargavan2016downgrade}, applied one layer lower, in the combiner itself.

\paragraph{When binding is necessary versus defence-in-depth.} When a hybrid KEM is embedded in TLS~1.3, IKEv2, or EDHOC, the protocol already authenticates the transcript and provides downgrade resilience; combiner-layer binding is then defence-in-depth. Binding is \emph{necessary} precisely when the embedding provides no such guarantee: a standalone drop-in KEM used outside a transcript-authenticating protocol, a bespoke raw-KEM pairing in a custom constrained handshake, or an early hybrid prototype. In those settings, binding makes downgrade resilience a self-contained property of the combiner, independent of whether the surrounding protocol is present and correctly configured, at the cost of one hash.

\paragraph{Relation to X-Wing.} X-Wing~\cite{xwing2024} and this construction answer different questions and are not in competition. X-Wing is a fixed combined KEM with no notion of negotiation; it is secure by construction when its inputs are fixed and is intended as a drop-in replacement for a single KEM. The present combiner adds the negotiation transcript to the key schedule so that the choice of suite, not only the shared secrets, is authenticated. A deployment that uses X-Wing inside a protocol with its own transcript binding needs nothing further; one that performs in-band suite selection without that binding gains exactly the missing guarantee from Equation~\eqref{eq:bound}.

\paragraph{Limitations.} The strongest-link bound is proven in the classical random-oracle model; the QROM guarantee is qualitative, with only a sketch in Appendix~\ref{secA2}. The device figures are a calibrated model composed from published per-primitive Cortex-M4F measurements from two campaigns, not a single instrumented deployment. The construction authenticates the integrity of the negotiated parameters and assumes entity authentication is provided separately; without it, a man-in-the-middle running two independent handshakes is outside the model. Finally, transcript binding captures negotiation integrity but not availability: an attacker can still force an abort by corrupting messages, which is a denial-of-service outcome, not a downgrade.

\paragraph{Threats to validity.} Internally, composing ML-KEM and X25519 cycle counts from two independent campaigns on comparable cores means absolute figures carry that caveat, although the relative comparisons are platform-independent. Externally, the latency analysis projects computation onto published round-trip datasets rather than measuring a live deployment, and the energy operating points represent a class of 802.15.4-class radios rather than one device. Reproducing the end-to-end cost on a single instrumented board is the highest-value next step.

\section{Conclusion}\label{sec:conc}

Hybrid key establishment secures the derived session key under the assumption that one component primitive survives, but it does not secure the negotiation that decides which primitives are used. A hybrid KEM proven secure as a fixed object gives no guarantee against an attacker who strips the post-quantum option, particularly when the combiner is deployed outside a protocol that authenticates the transcript. We defined downgrade resilience at the combiner layer, proved a two-sided separation between the plain and transcript-bound key schedules, and gave an explicit strongest-link bound corroborated by a distinguisher that tracks its analytic term. On a calibrated Cortex-M4 cost model with a radio-inclusive energy account, the defence costs one transcript hash: 11.8\% of compute and 1.5\% of total handshake energy, with no additional messages or bytes. For the constrained post-quantum deployments where protocol-level transcript authentication is absent or unreliable, this is a minimal, local, and self-contained way to close the downgrade gap.

Future work includes deriving a tight QROM constant for the transcript-bound combiner, extending the model to multi-session settings with session-state reveal, applying the principle to three-or-more component combiners for post-quantum agility, replacing the analytic energy model with power-instrumented single-board measurements, and integrating the combiner into a complete EDHOC extension evaluated against the existing cipher-suite negotiation.

\backmatter

\bmhead{Supplementary information}
The reference implementation, all experiment scripts, generated result files, figure-generation code, and the 30-check validation gate are provided as an accompanying software artifact.

\bmhead{Acknowledgements}
Not applicable.

\section*{Declarations}

\bmhead{Funding}
No funding was received for conducting this study.

\bmhead{Competing interests}
The authors have no competing interests to declare that are relevant to the content of this article.

\bmhead{Ethics approval and consent to participate}
Not applicable.

\bmhead{Consent for publication}
Not applicable.

\bmhead{Data availability}
The benchmarking inputs are derived from the publicly available pqm4 Cortex-M4 measurements and published round-trip latency datasets. No new empirical datasets were generated.

\bmhead{Materials availability}
Not applicable.

\bmhead{Code availability}
The protocol harness, experiment scripts, cost model, figure-generation code, and validation gate are available from the authors and will be released in a public repository upon acceptance.

\bmhead{Author contribution}
B.\ Gupta developed the construction, proofs, implementation, and experiments, and wrote the manuscript. S.\ Rana supervised the work and revised the manuscript. Both authors reviewed and approved the final version.

\bmhead{Use of AI tools}
An AI-based tool was used for language copy-editing (readability and grammar) only. All definitions, theorems, proofs, implementation, and reported numbers were produced and verified by the authors; no results were generated by an AI tool.

\begin{appendices}

\section{Full Proofs}\label{secA1}

\paragraph{Proof of Lemma~\ref{lem:correct}.} On an honest, unmodified hybrid handshake, $I$ and $R$ observe the same message sequence, so $\tau_I=\tau_R$. ML-KEM-768 decapsulation returns the encapsulated $\Kpq$ except with probability $\delta_{\mathrm{MLKEM}}\approx 2^{-164}$~\cite{fips203}, and X25519 yields equal $\Kec$ deterministically. Both $\KDF$ inputs coincide, so $K_I=K_R$, and the tag check $\MAC(K_I,\tau_I)=t_R$ holds. The only failure mode is ML-KEM decapsulation failure. \hfill$\square$

\paragraph{Proof of Theorem~\ref{thm:strong}.} Game~0 is the standard IND-CCA game for the hybrid KEM. Let $\mathsf{Query}$ be the event that $\mathcal{A}$ queries $\KDF$ at the challenge point $\ell\concat\Kpq\concat\Kec\concat\Hash(\tau)$. In Game~1 the challenger samples the session key uniformly and independently of the challenge ciphertext; Games~0 and~1 are identical unless $\mathsf{Query}$ occurs, so $\Adv_0\le\Adv_1+\Pr[\mathsf{Query}]$, and since the Game-1 key is independent of the challenge bit, $\Adv_1=0$.

Reduction $\mathcal{B}_{\mathrm{pq}}$ receives an ML-KEM-768 IND-CCA challenge $(ek^\ast,c^\ast,K^\ast_b)$, plants it as the post-quantum component, answers hybrid decapsulation queries with its own ML-KEM oracle for all $c\neq c^\ast$, and monitors $\mathcal{A}$'s $\KDF$ queries. If any query contains the challenge $\Kpq^\ast$ in the correct field, $\mathcal{B}_{\mathrm{pq}}$ distinguishes $K^\ast_b$. Hence $\Pr[\mathsf{Query}\wedge\Kpq\text{ unknown}]\le\Adv_{\mathrm{pq}}$; symmetrically $\Pr[\mathsf{Query}\wedge\Kec\text{ unknown}]\le\Adv_{\mathrm{ec}}$. The remaining $q_H/2^{\gamma}$ term bounds the probability of guessing an unqueried $\gamma$-bit secret. Taking the $\min$ over both reductions gives the bound. \hfill$\square$

\paragraph{Proof of Theorem~\ref{thm:dr}.} (i) The adversary intercepts $A_I$ before it reaches $R$ and removes every post-quantum entry, forwarding $A_I'=\{\textsf{x25519}\}$. $R$ selects \textsf{x25519}, computes $\Kec$, and sets $\Kpq=\varepsilon$. The plain key $K=\KDF(\Kpq\concat\Kec)$ depends only on $\Kec$, not on $A_I$, $A_I'$, or the selected suite. $I$ computes $K$ from the same $\Kec$, both parties accept on \textsf{x25519}, and the weaker-suite event holds with probability~1.

(ii) A successful downgrade requires an honest session to accept on a transcript $\tau_I\neq\tau_R$ (since $\mathcal{A}$ altered the negotiation) while $\MAC(K_I,\tau_I)=t_R$, where $t_R=\MAC(K_R,\tau_R)$ and $K_R=\KDF(\ell\concat\Kpq\concat\Kec\concat\Hash(\tau_R))$. If $\Hash(\tau_I)=\Hash(\tau_R)$, then $\mathcal{A}$ found a hash collision, with probability at most $q_H/2^n$ in the random-oracle model. Otherwise $K_I\neq K_R$ (distinct random-oracle inputs), and matching a tag computed under the independent $K_R$ requires an EUF-CMA forgery, with probability at most $\Adv^{\mathrm{euf\text{-}cma}}_{\MAC}$. Summing gives the bound. \hfill$\square$

\paragraph{Proof of Proposition~\ref{prop:cost}.} The two combiners share one protocol skeleton. The bound variant adds one $\Hash$ evaluation over $\tau$ per party before the $\KDF$ call and uses $\tau$ rather than a constant as the $\MAC$ input. No key generation, encapsulation, decapsulation, message, round trip, or wire byte is added. \hfill$\square$

\section{Quantum Random-Oracle Sketch}\label{secA2}

We sketch why the strongest-link guarantee survives a quantum adversary that queries $\Hash$ and $\KDF$ in superposition; a tight constant is left to future work. The key step in Theorem~\ref{thm:strong} is bounding $\Pr[\mathsf{Query}]$, the probability that the adversary evaluates $\KDF$ at the challenge point containing a component secret it does not otherwise recover. In the QROM this probability is bounded using the one-way-to-hiding (O2H) framework and its semi-classical-oracle refinement~\cite{unruh2015o2h,ambainis2019o2h}: the difference between the real and the reprogrammed oracle is bounded by the square root of the probability that a measured query hits the reprogrammed point. Instantiating the O2H bound with the guessing probability $q_H/2^{\gamma}$ replaces that term by $O(q_H/2^{\gamma/2})$, the expected Grover-style degradation, and leaves the component-KEM reductions intact because ML-KEM's own IND-CCA analysis is already QROM-sound~\cite{hofheinz2017fo,bos2018kyber}. Bindel et al.~\cite{bindel2019hybrid} carry out the analogous lifting for hash-based hybrid combiners in full; our combiner is of that form. For downgrade resilience (Theorem~\ref{thm:dr}(ii)), the collision term $q_H/2^n$ becomes $O(q_H^{3}/2^{n})$ under the standard quantum collision bound, which for $n=256$ remains below $2^{-100}$ for any realistic $q_H$.

\section{Validation-Gate Categories}\label{secA3}

Table~\ref{tab:gate} itemizes the 30-check validation gate that must pass before any quantitative result is reported. In the released run all 30 checks pass.

\begin{table}[t]
\caption{Validation-gate categories. All checks pass in the released run}\label{tab:gate}
\begin{tabular}{@{}llr@{}}
\toprule
Category & What it verifies & Checks \\
\midrule
Correctness   & honest handshakes agree; key length; suite selection & 6 \\
Downgrade     & plain downgraded, bound aborts, no false abort        & 8 \\
Distinguisher & fit slope vs analytic, $R^2$, monotonicity, CIs      & 6 \\
Cost model    & binding increment, wire-byte parity, additivity       & 4 \\
Known-answer  & SHA3-256 KAT, uniformity, determinism, sensitivity    & 4 \\
Footprint     & peak stack within budget, wire size bound             & 2 \\
\midrule
Total         &                                                       & 30 \\
\botrule
\end{tabular}
\end{table}

\section{Ablation: Binding Overhead versus Negotiation Size}\label{secA4}

\begin{figure}[t]
\centering
\includegraphics[width=\columnwidth]{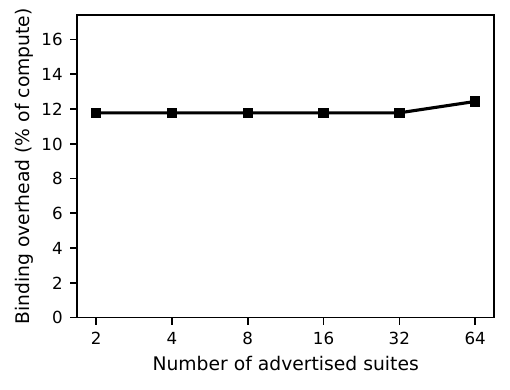}
\caption{Transcript-binding overhead as a fraction of handshake compute, versus the number of advertised suites. The overhead is dominated by a single short Keccak hash and remains near 12\%}\label{fig:ablation}
\end{figure}

Figure~\ref{fig:ablation} shows that transcript-binding overhead, as a fraction of total handshake compute, is essentially flat as the number of advertised suites grows. The transcript hash spans the fixed public-key and ciphertext material plus a few identifier bytes per advertised suite, so the number of Keccak-$f$ permutations grows slowly, and the ML-KEM-768 payload always dominates. The overhead stays near 12\% regardless of negotiation complexity, so the defence does not become the bottleneck as negotiations grow.

\end{appendices}

\bibliography{refs}

\end{document}